\documentclass[aps,pra,superscriptaddress,twocolumn,longbibliography]{revtex4-1}
\usepackage{units}
\usepackage{amsmath}
\usepackage{amsthm}
\usepackage{amssymb}
\usepackage{graphicx}
\usepackage{color}
\usepackage{bbold}
\usepackage[colorlinks=true, urlcolor=blue, citecolor=blue,linkcolor=blue,citebordercolor={1 0 0},linkbordercolor={0 0 1}]{hyperref}

\theoremstyle{plain}
\newtheorem{thm}{\protect\theoremname}
\theoremstyle{plain}
\newtheorem{fact}[thm]{\protect\factname}
\ifx\proof\undefined
\newenvironment{proof}[1][\protect\proofname]{\par
	\normalfont\topsep6\p@\@plus6\p@\relax
	\trivlist
	\itemindent\parindent
	\item[\hskip\labelsep\scshape #1]\ignorespaces
}{%
	\endtrivlist\@endpefalse
}
\providecommand{\proofname}{Proof}
\fi
\theoremstyle{plain}
\newtheorem{lem}[thm]{\protect\lemmaname}
\theoremstyle{remark}

\newcommand{\bra}[1]{\langle #1|}
\newcommand{\ket}[1]{|#1 \rangle}
\makeatother
\newcommand{\braket}[2]{\langle #1 \vert #2 \rangle}
\newcommand{\idg}[1]{{\bfseries #1)}}

\newcommand\numberthis{\addtocounter{equation}{1}\tag{\theequation}}
\providecommand{\factname}{Fact}
\providecommand{\theoremname}{Theorem}
\providecommand{\claimname}{Claim}
\providecommand{\lemmaname}{Lemma}
\providecommand{\definitionname}{Definition}

\definecolor{KB}{rgb}{0.4,0.3,0.9}

\definecolor{THc}{rgb}{0.9,0.3,0.2}

\newcommand{\revA}[1]{{#1}}

\theoremstyle{definition}
\newtheorem{defn}[thm]{\protect\definitionname}

\graphicspath{{./Images/},{./ImagesAppendix/}}

\newcommand{\subfigimg}[3][,]{%
	\setbox1=\hbox{\includegraphics[#1]{#3}}
	\leavevmode\rlap{\usebox1}
	\rlap{\hspace*{2pt}\raisebox{\dimexpr\ht1-0.5\baselineskip}{{\bfseries \large\textsf{#2}}}}
	\phantom{\usebox1}
}

\begin{document}
\title{Iterative Quantum Assisted Eigensolver}
\author{Kishor Bharti}
\email{kishor.bharti1@gmail.com}
\affiliation{Centre for Quantum Technologies, National University of Singapore 117543, Singapore}
\author{Tobias Haug}
\affiliation{Centre for Quantum Technologies, National University of Singapore 117543, Singapore}
\begin{abstract}
The task of estimating the ground state of Hamiltonians is an important problem in physics with numerous applications ranging from solid-state
physics to combinatorial optimization. We provide a hybrid quantum-classical algorithm for approximating the ground state of a Hamiltonian that builds on the powerful Krylov subspace method in a way that is suitable for current quantum computers. 
Our algorithm systematically constructs the Ansatz using any given choice of the initial
state and the unitaries describing the Hamiltonian. \revA{The only task of the quantum
computer is to measure overlaps and no feedback loops are required.}
The measurements can be performed efficiently on current quantum hardware without requiring any complicated measurements such as the Hadamard test.
Finally, a classical computer solves a well characterized quadratically constrained optimization program.
\revA{Our algorithm can reuse previous measurements to calculate the ground state of a wide range of Hamiltonians without requiring additional quantum resources. Further, we demonstrate our algorithm for solving problems consisting of thousands of qubits.} The algorithm works for almost every
random choice of the initial state and circumvents the barren plateau problem. 
\end{abstract}
\maketitle

\noindent {\em Introduction.---} The recent ``Quantum Supremacy'' experiment by researchers at Google \cite{arute2019quantum}
has demonstrated the ability of quantum devices to execute computation beyond the reach of any classical computing device. Despite the hope generated by the aforementioned experiment, the absence of quantum advantage for any practical use-case has been a subject of concern.
The search for a ``Killer app'' for noisy intermediate-scale quantum (NISQ) \cite{preskill2018quantum,bharti2021noisy} devices continue, with none available at the moment.

Some of the potential areas where a killer app could be useful are solid state physics, quantum chemistry and combinatorial optimization.
Many of these problems can
be recast as Hamiltonian ground state finding problem. Some of the
canonical NISQ era algorithms for approximating the ground state of Hamiltonians are variational quantum eigensolver
(VQE) \cite{peruzzo2014variational,mcclean2016theory,kandala2017hardware} and quantum approximate optimization algorithm (QAOA) \cite{farhi2014quantum,farhi2016quantum}. They employ a classical optimizer to adjust the parameters of
a parametric quantum circuit while utilizing a quantum device to calculate the objective function or its gradient. These algorithms have been regarded as one of the most promising candidates for near-term quantum advantage \cite{harrow2017quantum, farhi2016quantum,mcardle2020quantum}. The training parameter
landscape corresponding to these algorithms is highly non-convex and is an uncharacterized optimization program
in general, \revA{which renders systematic studies difficult~\cite{bittel2021training}.} 
Furthermore, the appearance of vanishing
gradient problem or barren plateau problem as the circuit size or hardware noise increases
has led to valid concern regarding the future of these algorithms~\cite{mcclean2018barren,huang2019near,sharma2020trainability,cerezo2020cost,wang2020noise}
Recently, many algorithms based on \revA{computing overlaps between quantum states}
have been suggested in the literature to estimate the ground state of Hamiltonians~\cite{mcclean2017hybrid,kyriienko2020quantum,parrish2019quantum,bespalova2020hamiltonian,huggins2020non,takeshita2020increasing,stair2020multireference,motta2020determining,seki2020quantum}. \revA{In this work, we propose 
a systematic and general approach to create a NISQ compatible, sufficiently expressible and trainable Ansatz. Our ansatz does not require complicated measurement methods such as the Hadamard test (see section ~\ref{sec: measurements} in the Supplementary Materials (SM)). For a comparison of the current work with existing matrix overlap computation based variational algorithms, refer to section~\ref{sec: compexist} in SM.}

Recently, the quantum assisted eigensolver (QAE) was proposed~\cite{bharti2020quantum}, a hybrid quantum-classical algorithm based on overlap matrix computation to approximate the ground state of a Hamiltonian. \revA{The classical part of QAE is a quadratically constrained quadratic program (QCQP) which is  a well characterized optimization program with efficiently computable lower bounds. A sufficient condition for a local minimum to be a global minimum is also provided by the QAE algorithm.  A solver can employ the aforementioned condition as a stopping criterion.} Despite its rich mathematical structure, QAE does not provide a strategy to generate an expressible Ansatz for a given problem Hamiltonian and to measure overlaps in a NISQ-friendly way.

In this Letter, we present a novel hybrid classical-quantum algorithm for
approximating the ground state and ground state energy of a Hamiltonian, called iterative quantum assisted eigensolver (IQAE).
Without losing generality, the Hamiltonian is assumed to be a linear combination
of unitaries. For a given choice of the initial state and the
unitaries describing the Hamiltonian, our algorithm
iteratively builds the Ansatz using the unitaries. The only job of the quantum computer
is to compute overlap matricess, which can be performed efficiently on the existing quantum hardware without requiring any complicated measurement \revA{or feedback loops}. At
the end of a given iteration, the classical computer solves a QCQP. If a sufficient accuray has been reached, we stop the algorithm or else proceed to the next iteration.

\begin{figure}[htbp]
	\centering
	\subfigimg[width=0.5\textwidth]{}{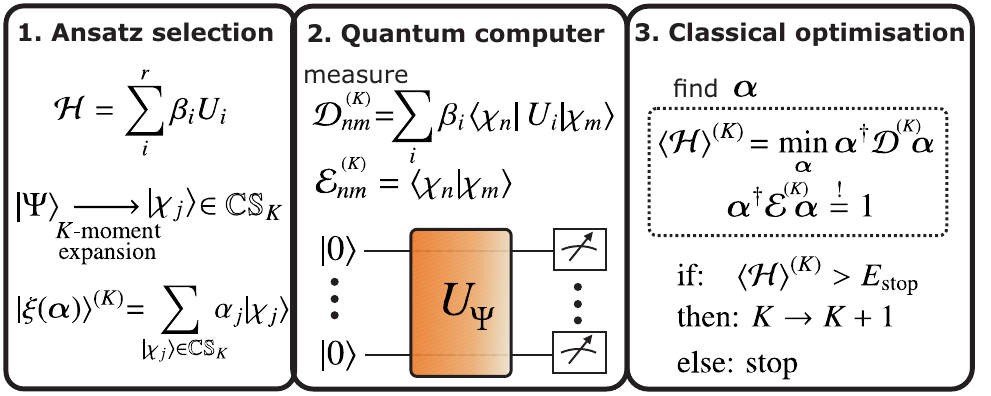}
	\caption{The IQAE algorithm systematically builds the Ansatz using some efficiently implementable quantum state and the unitaries which describe the Hamiltonian. The Ansatz is represented as a linear combination of the fine-grained Krylov subspace basis or cumulative $K$-moment states which are inspired by the Krylov subspace method (see Definition \ref{def:cumulant_states}). The second step computes overlap matrices on a quantum computer. Finally, the optimization program \eqref{eq:P1} is executed on a classical computer. The approximation of the ground state improves with increasing $K$.}
	\label{Fig: IQAE}
\end{figure}

\medskip
\noindent {\em Background.---} We start with a  synopsis of the QAE algorithm \cite{bharti2020quantum}. Let us consider
the problem Hamiltonian $H$ to be a linear combination of $N$-qubit unitaries,
\begin{equation}
H=\sum_{i=1}^{r}\beta_{i}U_{i}\, ,\label{eq:ham_unitaries}
\end{equation}
where the combination coefficients $\beta_{i}\in\mathbb{C}$ and the
unitaries $U_{i}\in SU\left(2^{N}\equiv \mathcal{N}\right)$, for $i\in\left\{ 1,2,\cdots,r\right\} $. 
Our goal is to provide an approximation to the ground state $\left(\vert\phi_{1}\rangle\right)$ and the
ground state energy $\lambda_{1}$ of $H$.

In step $1$ of the QAE algorithm
one selects a set of $L$ quantum states $\left\{ \vert\chi_{i}\rangle\right\} _{i=1}^{L}$
and builds the Ansatz as 
\begin{equation}
\vert\xi\left(\boldsymbol{\alpha}\right)\rangle=\sum_{i=1}^{L}\alpha_{i}\vert\chi_{i}\rangle,\label{eq:Ansatz}
\end{equation}
for $\boldsymbol{\alpha}\in\mathbb{C}^{L}.$ Then, one finds a suitable $\boldsymbol{\alpha}$
which minimizes the expectation value of the Hamiltonian in Eq.\ref{eq:ham_unitaries} with the following optimization program
\begin{equation}
\text{minimize }\boldsymbol{\alpha}^{T}\mathcal{D}\boldsymbol{\alpha}\,\text{ subject to }\boldsymbol{\alpha}^{T}\mathcal{E}\boldsymbol{\alpha}=1.\label{eq:P1}
\end{equation}
In the above expression the $(n,m)$th element of the $L\times L$
matrices $\mathcal{D}$ and $\mathcal{E}$ are given by
\begin{align}
\mathcal{D}_{n,m}&=\sum_{i}\beta_{i}\langle\chi_{n}\vert U_{i}\vert\chi_{m}\rangle\label{eq:D_matrix}\\
\mathcal{E}_{n,m}&=\langle\chi_{n}\vert\chi_{m}\rangle,\label{eq:E_matrix}
\end{align}
for all possible valid $(n,m).$ 
Step $2$ of the QAE algorithm
is the calculation of the overlap matrices $\mathcal{D}$ and $\mathcal{E}$ on a quantum
computer, which concludes
the job of the quantum computer for this iteration. 

The final step $3$ of the QAE algorithm is post-processing the measured overlaps with optimization program \eqref{eq:P1} on a classical computer. \revA{Note that the optimization program  \eqref{eq:P1} is a QCQP, which is a well-studied optimization program in the classical optimization literature ~\cite{Boyd2004}.} Alternatively, one can follow the classic Rayleigh–Ritz procedure, and approximate the ground state by solving for the smallest eigenvalue $\lambda$ of the generalized eigenvalue problem $\mathcal{D}\boldsymbol{\alpha}=\lambda\mathcal{E}\boldsymbol{\alpha}$~\cite{jia2001analysis,mcclean2017hybrid}.
The choice of Ansatz in step $1$ of the QAE algorithm heavily determines the proximity of the final output to the correct value, and hence systematic approaches should be developed to select the Ansatz. Moreover, step $2$ of the QAE algorithm involves the computation of overlap matrices $\mathcal{D}$ and $\mathcal{E}$ which in general requires the Hadamard test implemented with controlled unitaries. Since implementing them is challenging, one needs to devise alternative approaches.

\revA{\medskip
\noindent {\em Krylov subspace based algorithm.---} 
We now introduce Krylov subspace based algorithm that powers our IQAE algorithm. Given a scalar $\tau$, an $\mathcal{N}\times \mathcal{N}$ matrix $A$ and
a vector $v$ of length $\mathcal{N}$, the matrix exponential operator
$\exp\left(-\tau A\right)$ applied on $v$ can be approximated as $\exp\left(-\tau A\right)v\approx p_{K}\left(-\tau A\right)v$
where $p_{K}$ is a degree $K$ polynomial~\cite{lanczos1950iteration,saad1992analysis,seki2020quantum,motta2020determining}. This approximation
is an element of the Krylov subspace
\begin{equation}
Kr_{K}\equiv span\left\{ v,Av,\cdots,A^{K}v\right\} .\label{eq:Krylov_1}
\end{equation}
Thus, the problem of approximating $\exp\left(-\tau A\right)v$ can
be recast as finding an element from $Kr_{K}$, which becomes exact for $K-1=rank(A).$
In our case, we can identify $v$ with the initial state $\vert\psi\rangle$
and $A$ with the Hamiltonian $H$. Let us consider the 
Ansatz
\begin{equation}
\vert\varsigma\left(\alpha\right)\rangle^{\left(K\right)}=\alpha_{0}\vert\psi\rangle+\alpha_{1}H\vert\psi\rangle+\cdots+\alpha_{K}H^{K}\vert\psi\rangle,\label{eq:Krylov_Ansatz}
\end{equation}
where $\alpha_{i}\in\mathbb{C}$ for $i\in\left\{ 0,1,\ldots,K\right\}.$
The task of approximating ground state can be trivially recast as a QCQP (see section \ref{sec: Krylov} in SM).
We can relate Eq.\eqref{eq:Krylov_Ansatz} directly to imaginary time evolution, which is given by
\begin{equation}
\vert\varsigma(\tau)\rangle=\frac{1}{\gamma} e^{-\tau H}\vert\psi\rangle=\frac{1}{\gamma}\sum_{p=0}^{\infty}\frac{\left(-\tau H\right)^{p}}{p!}\vert\psi\rangle
\end{equation}
where $\gamma$ is normalizing the quantum state.  If $\vert\psi\rangle$ has non-zero overlap with the ground state, then in the limit of $\tau\rightarrow\infty$ the ground state $\ket{\phi_1}$ of $H$ is $\vert\varsigma(\tau\rightarrow\infty)\rangle\rightarrow\ket{\phi_1}$. We can think of the Krylov subspace approach as a truncated version of imaginary time evolution (see section \ref{sec: justify} in SM for additional details). However, imaginary time evolution is a non-unitary operation that is highly challenging to implement on quantum computers which are unitary by nature~\cite{motta2020determining}.
When implementing the Krylov subspace ansatz with QAE, we have to estimate overlaps involving $H^k$, which is a challenge for NISQ computers. Importance sampling has been proposed to estimate the expectation value of  $H^k$~\cite{mcclean2020decoding}, however this may require more measurements than current NISQ computers can handle for sufficient accuracy.

\noindent {\em IQAE algorithm.---}
We now propose the IQAE algorithm that provides a systematic
approach to select the Ansatz via the Krylov subspace idea and an efficient estimation of the overlaps.
We assume that our Hamiltonian is a sum of unitaries $H=\sum_i\beta_i U_i$ and build the Krylov subspace $Kr_{K}\equiv span\left\{ \ket{\psi},H\ket{\psi},\cdots,H^{K}\ket{\psi}\right\}$ up to order $K$ using state $\ket{\psi}$.
We take each element of the subspace $H^k\ket{\psi}=(\sum_i\beta_i U_i)^k\ket{\psi}$ and multiply out the power $k$ such that we get a sum. Each constituent term of the sum $\prod_{{i_1},\dots,{i_k}}U_{i_{1}}\dots U_{i_k}\ket{\psi}$ is then added to a set $\mathbb{S}_k$. Finally, we combine all $K+1$ sets $\mathbb{S}_k$ into what we now call the fine-grained Krylov subspace basis or cumulative $K$-moment states $\mathbb{CS}_{K}\equiv\cup_{j=0}^{K}\mathbb{S}_{j}$, which is formally defined below.
\begin{defn}  \label{def:cumulant_states}
Given a set of unitaries $\mathbb{U}\equiv\left\{ U_{i}\right\}_{i=1}^{r}$, a positive integer $K $ and some quantum state $\vert\psi\rangle,$ $K$-moment states is the set of quantum states of the form $\left\{ U_{i_{K}}\cdots U_{i_{2}}U_{i_{1}}\vert\psi\rangle\right\} _{i}$ for  $U_{i_{l}}\in\mathbb{U}.$ We denote the aforementioned set by $\mathbb{S}_{K}$. The fine-grained Krylov subspace basis or cumulative $K$-moment states $\mathbb{CS}_{K}$ is defined as $\mathbb{CS}_{K}\equiv\cup_{j=0}^{K}\mathbb{S}_{j}$.
\end{defn}
In the following, we use $\mathbb{CS}_K$ as ansatz for IQAE
\begin{equation}
\vert\xi\left(\boldsymbol{\alpha}\right)\rangle^{\left(K\right)}=\sum_{\vert \chi_{i}\rangle\in\mathbb{CS}_{K}}\alpha_{i}\vert \chi_{i}\rangle\,.\label{eq:cumulative_moment_j_Ansatz}
\end{equation}
For example, the fine-grained Krylov subspace basis for order $K$ is given by $\mathbb{CS}_{K}=\{\vert \psi \rangle\} \cup \left\{ U_{i_1}\vert\psi\rangle\right\} _{i_1=1}^{r} \cup \dots \cup \left\{ U_{i_K}\dots U_{i_1}\vert\psi\rangle\right\} _{i_1=1,\dots,i_K=1}^{r}$.

Then, we measure the overlaps of $\mathcal{D}$ and $\mathcal{E}$.
When we assume that each unitary $U_i$ of the Hamiltonian acts only on a logarithmic number of qubits $\mathcal{O}\left(poly\left(logN\right)\right)$ qubits, we can use direct methods~\cite{mitarai2019methodology} to efficiently  estimate the overlaps without requiring the Hadamard test. The efficiency in the aforementioned line is in terms of the number of gates and classical post-processing time. For details refer to Fact~\ref{fact: hadamard} in SM. Note that the approach based on Fact~\ref{fact: hadamard} will be infeasible for an Ansatz with higher values of $K$.
If the unitaries $U_i$ in Eq.\eqref{eq:ham_unitaries} are tensored Pauli matrices or Pauli strings of the form $P_i=\bigotimes_{j=1}^N\boldsymbol{\sigma}_j$, with $\boldsymbol{\sigma}_j\in \{\mathbb{1},\sigma^x,\sigma^y,\sigma^z\}$, the measurement of the overlaps can be further simplified.
The matrix elements of $\mathcal{D}$ and $\mathcal{E}$ are then of the form $\bra{\psi}\prod_{q} P_{q}\ket{\psi}=a\bra{\psi}P'\ket{\psi}$, which can be rewritten as a single Pauli string $P'$ with a prefactor $a\in\{1,-1,i,-i\}$ (see SM for derivation). Thus, the matrix elements can be efficiently measured on NISQ devices by sampling in a Pauli rotated basis for any number of qubits.
Using the overlaps, we run the classical post-processing step with the QCQP to find the weights $\boldsymbol{\alpha}$ that approximate the ground state.
We can estimate the rate of converge with the difference of estimated ground state energy between order $K-1$ and $K$ of IQAE $\Delta E_K=E_{K}-E_{K-1}$. If $\Delta E<E_\text{c}$ is smaller than a chosen threshold $E_\text{c}$, the algorithm halts, else we set $K\rightarrow K+1$ and run IQAE with higher order terms of the Krylov subspace. Alternatively, we may also use a classical calculable estimate of the ground state energy as a stopping criterion.
We note that our approach is more powerful than the original Krylov subspace method or imaginary time evolution. Both methods only converge if the state $\ket{\psi}$ has a non-zero overlap with the groundstate $\braket{\psi}{\phi_1}\neq0$. For random initial states, this condition is fulfilled with high probability, but for Hamiltonians with symmetries, $\ket{\psi}$ may lie in a different symmetry sector than the groundstate. As imaginary time evolution or $H^k$ does not break the symmetry, it will not converge to the correct ground state. 
While our Ansatz is inspired by the Krylov subspace and imaginary time evolution, IQAE does not have this drawback in general. We demonstrate this with a toy model in  section~\ref{sec:imaginary} in SM, where the Hamiltonian $H=\sum_i\beta_i U_i$ has a particular symmetry while the Ansatz $\ket{\psi}$ is in a different symmetry sector than the ground state. As the individual unitaries $U_i$ break this symmetry, IQAE converges whereas imaginary time evolution does not.

\medskip
\noindent {\em Resource estimation.---} The number of terms for the ansatz in Eq.~\eqref{eq:Krylov_Ansatz} to exactly converge scales linearly with the rank of the Hamiltonian. As discussed earlier, due to the inability of the current term NISQ devices to calculate the overlap matrix elements corresponding to the aforementioned ansatz, we employ the $\mathbb{CS}_{K}$ ansatz. 
For a Hamiltonian $H$ of rank $p$, consisting of $r$ unitaries,
the size of the overlap matrices $\mathcal{D}$ and $\mathcal{E}$
for achieving perfect accuracy is upper bounded by $r^{2p}$. However, depending on the algebra of the unitaries which define the Hamiltonian, this number can be much lower as we demonstrate in the next section. For example, this is the case when the Hamiltonian is a linear combination of a small number of unitaries, the Hamiltonian is a low rank matrix or the Krylov subspace closes for a low order $K$ such for quantum many-body scars~\cite{serbyn2021quantum}. 
The accuracy of IQAE depends on the number of Ansatz states and the choice of the zero moment state. However, finding the ground state is a QMA-hard problem and thus any ground state solver will require in general an exponential resources on a quantum device. The individual elements of the overlap matrices can be calculated with additive accuracy $\epsilon$ and failure probability at most $\delta$ using $\mathcal{O}\left(\frac{1}{\epsilon^{2}}\log\left(\frac{1}{\delta}\right)\right)$ copies of $\vert\psi\rangle$ (see section ~\ref{sec: measurements} for details) in SM.}

\medskip
\noindent {\em Examples.---}
\begin{figure*}[htbp]
	\centering
		\subfigimg[width=0.26\textwidth]{a}{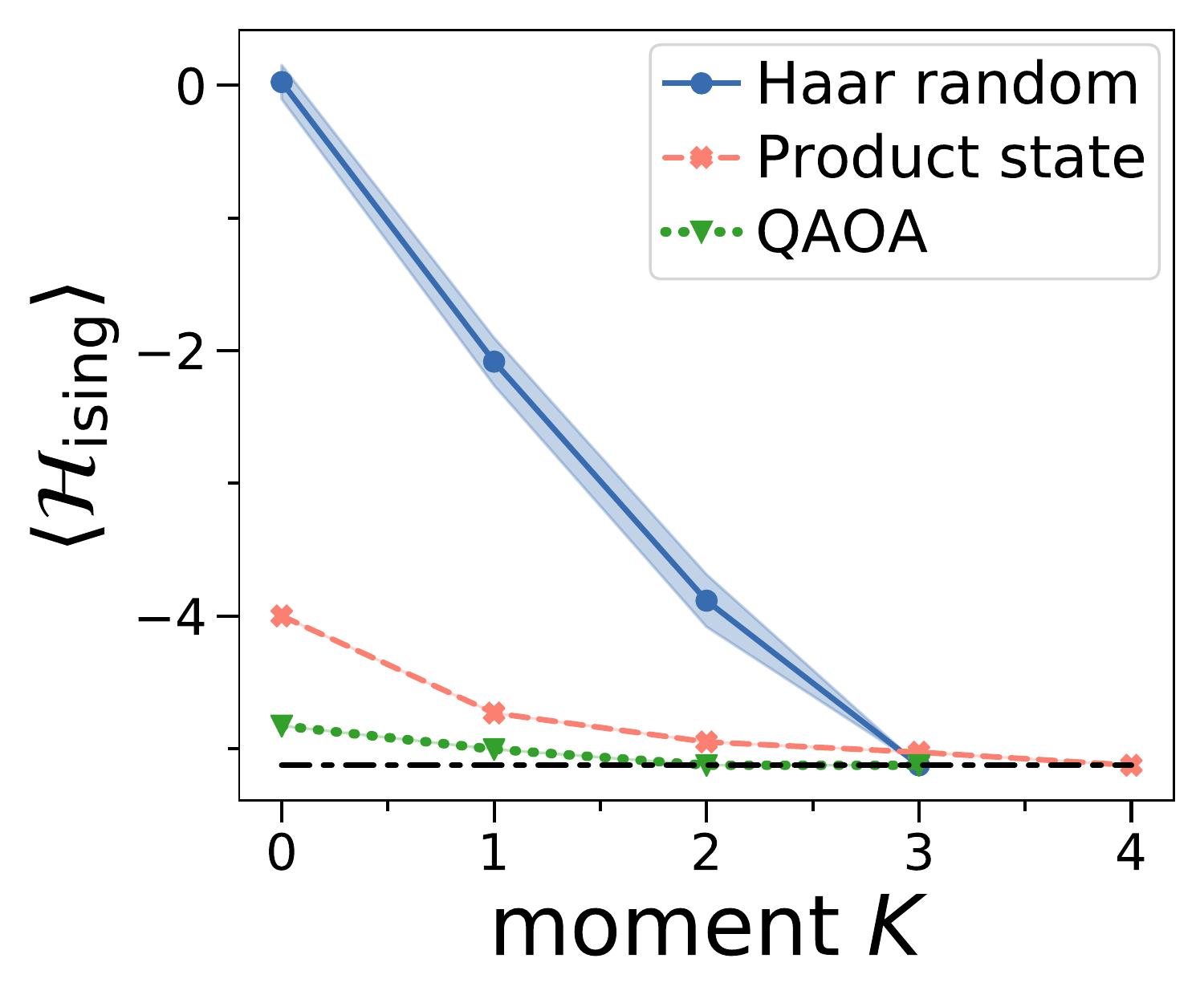}
	\subfigimg[width=0.26\textwidth]{b}{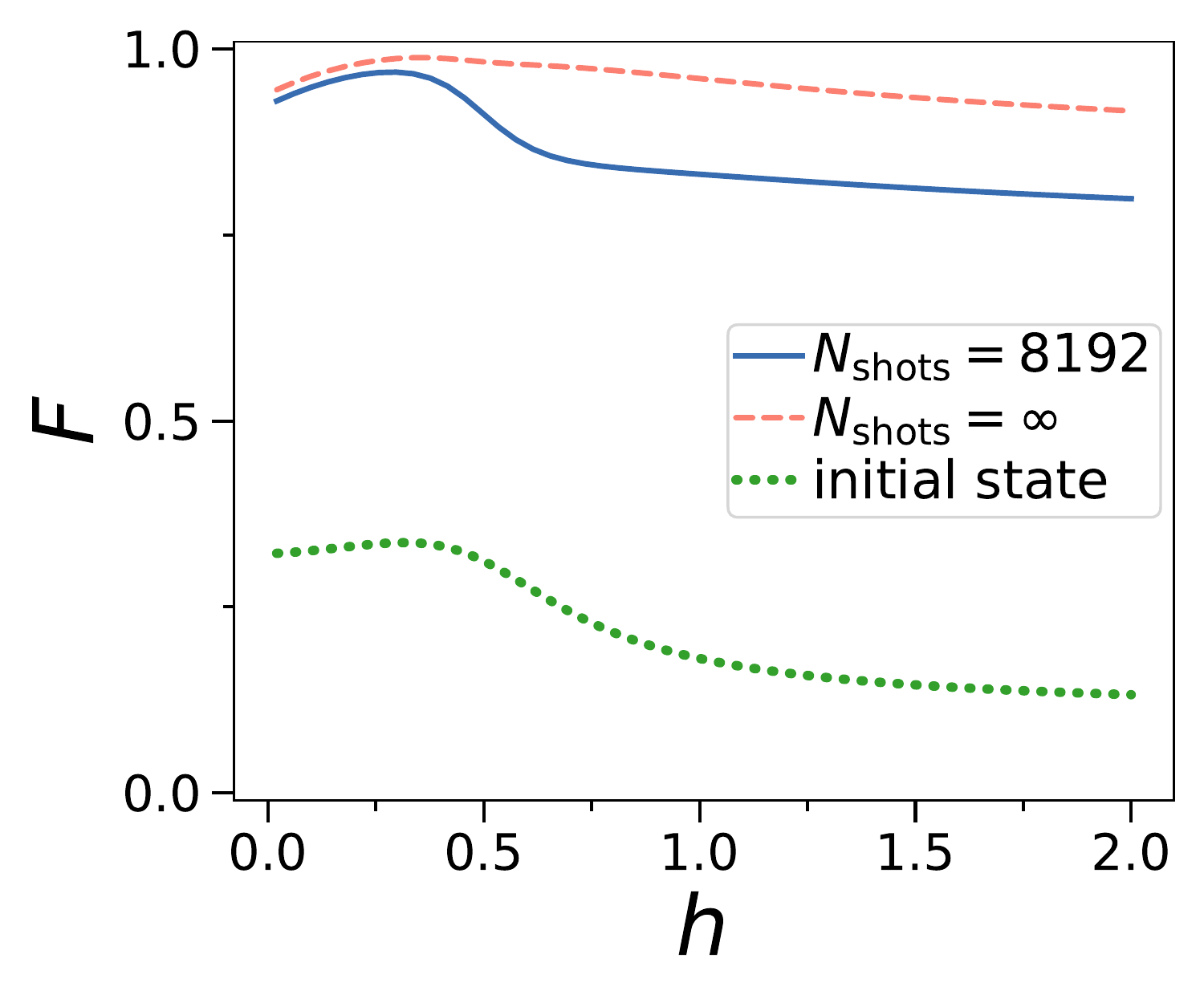}
	\subfigimg[width=0.26\textwidth]{c}{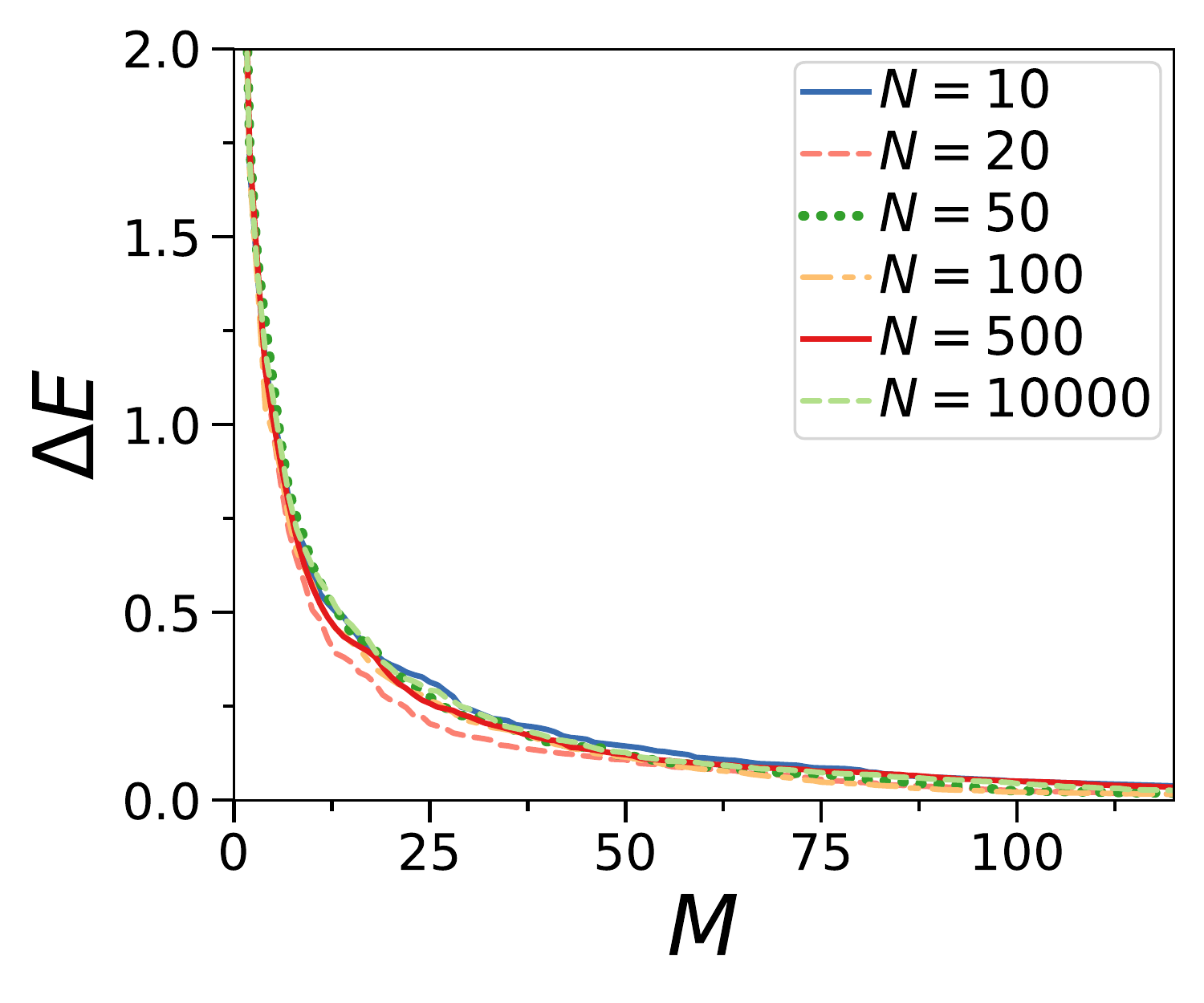}
	\caption{\idg{a} IQAE cost function plotted against order $K$ of the fine-grained Krylov subspace for the transverse Ising model with $N=8$ qubits and $h=\frac{1}{2}$. As ansatz, we use a random variational circuit with 200 layers, a product state and a QAOA ansatz  of depth $p=1$ $\prod_{k=1}^p\exp(-i\beta_k\sum_j\sigma^x_j)\exp(-i\gamma_kH)\ket{+}^{\otimes N}$ with pre-optimized $\gamma, \beta$ parameters. Shaded lines indicate standard deviation of the sampled cost function.
	\idg{b} Learning ground state of Ising Hamiltonian for all magnetic fields $h$ with a single run of IQAE. We plot fidelity $F=\vert{\braket{\phi_1}{\psi_\text{IQAE}}}\vert^2$ with the exact ground state against transverse field strength $h$. From the cumulative $K$-moment states, we select a subset of 100 quantum states for our ansatz. Then, we run IQAE once and use the measured overlaps to determine the ground state for all values of $h$, as $h$ can be varied for free during post-processing.
	We show the fidelity with the exact ground state for finite number of measurement shots $N_\text{shots}=8192$ and an infinite $N_\text{shots}=\infty$. The ansatz is a optimized $p=2$ QAOA ansatz for $h=\frac{1}{2}$. To evaluate the overlaps including the Hamiltonian expectation value, 6358 Pauli strings have to be calculated, which we measured by sampling 1098 different measurement settings.
	\idg{c} IQAE to calculate ground state of Hamiltonian consisting of $r=8$ random Pauli strings and many qubits $N$. We show difference between IQAE and exact ground state energy $\Delta E$ as function of number of ansatz states $M$. 
	}
	\label{fig:meta}
\end{figure*}
Expressive variational quantum circuits are known to suffer from the barren plateau problem, where the gradients of the circuit vanish with increasing number of qubits and render optimization challenging~\cite{mcclean2018barren}.
IQAE can train these types of circuits very easily, as we demonstrate for the Hamiltonian $H_\text{bp}=\sigma_1^z\sigma_2^z$ and a circuit $\ket{\psi}$ with $d\gg1$ layers, composed of alternating single-qubit rotations around randomly chosen $x$-, $y$- or $z$-axis parameterized with random angles and C-phase gates arranged in a hardware efficient manner. This circuit and Hamiltonian were first used to demonstrate the barren plateau problem for VQE~\cite{mcclean2018barren}. 
While this problem is difficult for VQE, IQAE converges already for the first moment $\mathbb{CS}_1=\left\{ \ket{\psi},\sigma_1^z\sigma_2^z\ket{\psi}\right\}$ as the Krylov subspace closes with next higher order moments remaining in the same subspace. 
\revA{Here, IQAE circumvents the barren plateau problem by avoiding to vary the parameters of quantum state and instead just uses it as a basis for the moment expansion.}

Next, we apply IQAE to a many-body problem, namely the transverse Ising model
$H_\text{ising}=\frac{1}{2}\sum_{i}\sigma^x_i\sigma^x_{i+1}-h\sum_{i}\sigma^z_i$.
In Fig.\ref{fig:meta}a, we investigate different types of Ans\"atze as function of the order $K$ of the fine-grained Krylov subspace basis. 

\revA{Next, we demonstrate that IQAE can calculate the ground state for various Hamiltonian parameters in a single iteration only. 
For IQAE, we measure the overlap elements $\bra{\psi_i} U_k\ket{\psi_j}$ of $\mathcal{D}_{i,j}=\sum_k\beta_k\bra{\psi_i} U_k\ket{\psi_j}$ (Eq.\ref{eq:D_matrix}). 
The weights $\beta_k$ are only used during the post-processing step and thus can be set to any value without requiring additional overlap measurements. Thus, after determining $\mathcal{D}$ and $\mathcal{E}$ once on the quantum computer, we can calculate the ground state of Hamiltonians for any choice of $\beta_k$.
In Fig.\ref{fig:meta}b, we use this feature to substantially reduce resources for calculating the ground state of the transverse Ising model for all values of $h$ by running IQAE only a single time. We restrict the set of $K$-moment states to 100 quantum states only and use only a budget of $9\cdot 10^6$ measurement shots in total or $1098$ measured quantum circuits by combining Pauli measurements using Clique covering~\cite{verteletskyi2020measurement}. Then, we can determine the ground state of the Ising model for all values of $h$ with high fidelity. The ansatz state is a optimised QAOA circuit of depth $p$. To account for fluctuations due to the shot noise, the overlap matrix $\mathcal{E}$ is regularized by setting all eigenvalues $\lambda$ of $\mathcal{E}$ that are smaller than $\lambda<10^{-1}$ to zero.}

\revA{In Fig.\ref{fig:meta}c, we solve a problem with an extensive number of qubits. The Hamiltonian are $r$ random Pauli strings $H=\sum_{i=1}^r \beta_i P_i$, where $P_i=\otimes_{j=1}^N \boldsymbol{\sigma}_j$ has randomly chosen $\boldsymbol{\sigma}_j\in\{I,\sigma^x,\sigma^y,\sigma^z\}$ and $\beta_i\in[-1,1]$.  
The initial state is the $N$-qubit product state with all zeros. We pick the first $M$ quantum states from the cumulative $K$-moment states to construct the ansatz. We find a fast convergence with increasing $M$ to the ground state energy for up to $N=10000$ qubits. The Krylov subspace closes for $K=r$, with maximal $M_\text{max}=2^r$ states.}

\begin{table*}[htbp]
\begin{minipage}{\textwidth}
\centering
\begin{tabular}{|c|c|c|}
\hline 
\textbf{Criteria} & \textbf{VQE} & \textbf{IQAE}\tabularnewline
\hline 
\hline 
Ansatz selection & Based on heuristic (in general) & Systematic\tabularnewline
\hline 
Optimization Program & Uncharacterized (in general)  & Quadratically constrained quadratic program (QCQP)\tabularnewline
\hline 
Random Initialization & Leads to barren plateau & Circumvents the barren plateau problem\tabularnewline
\hline 
Provable guarantees & Not possible in general & Mathematically rich to allow provable guarantees \cite{bharti2020quantum} \tabularnewline
\hline 
Feedback loop & Yes & No classical quantum feedback loop \cite{bharti2020quantum} \tabularnewline
\hline 
\end{tabular}
\caption{A comparison between VQE and IQAE}
\label{table: Comparison_VQE_IQAE}
\end{minipage}
\end{table*}
\medskip

\noindent {\em Discussion.---} We provided a hybrid quantum-classical algorithm to
approximate the ground state of a Hamiltonian.
The algorithm is iterative and systematically
builds the Ansatz using some efficiently preparable quantum state
and the unitaries which define the Hamiltonian. We construct our ansatz with imaginary time evolution and the Krylov subspace method in a NISQ-friendly way. 
The overlap matrices for our algorithm can be computed efficiently on current quantum computers using direct methods~\cite{mitarai2019methodology} without the need for complicated methods such as the Hadamard test (see section \ref{sec: measurements} in SM for details). 
If the Hamiltonian is a linear combination of Pauli strings, the elements can be directly inferred by sampling in the computational basis. The optimization program~\eqref{eq:P1} is a well characterized QCQP unlike variational quantum algorithms, which are difficult to characterize~\cite{bittel2021training}. 
\revA{Our algorithm can reuse previous measurement results to calculate the ground states of Hamiltonians $H=\sum_i \beta_iU_i$ for any choice of $\beta_i$ without requiring further quantum resources. This tremendously speeds up problems where one needs to vary the Hamiltonian parameters such as for the calculation of the bond length of molecules~\cite{cao2019quantum}. Further, we can solve the ground state of a Hamiltonian of multi-body Pauli strings for thousands of qubits. By using an entangled ansatz beyond classical simulability one could showcase the power of quantum computers. For other NISQ algorithms such as VQE it is challenging to handle many qubits and multi-body terms.}

In contrast to imaginary time evolution, the optimization of IQAE can proceed independently whether the initial state has finite overlap with the ground state as shown in section~\ref{sec: Numerical} in the SM.
While the classical optimization program of IQAE is non-convex in general, we can use convex relaxation or convex restriction to ensure efficient optimization~\cite{bharti2020quantum}.
We discuss possible future works in section~\ref{sec: future} in the SM.

Our algorithm is very general and subsumes VQE to enhance its scope (see section \ref{sec: QAE_VQE} in SM for details). We believe that our algorithm will
stimulate future research on classical techniques to approximate the
ground state of a Hamiltonian by harnessing the linear combination
of the unitary structure. As a final synopsis, we provide a brief comparison between our algorithm and VQE in Table \ref{table: Comparison_VQE_IQAE}.

\medskip
{\noindent {\em Acknowledgements---}} We are grateful to the
National Research Foundation and the Ministry of Education, Singapore
for financial support.
\bibliographystyle{apsrev4-1}
\bibliography{IQAE}

\appendix 
\section{Comparison with existing methods} \label{sec: compexist}
\revA{The general form of the ansatz used by the IQAE algorithm is
\begin{equation}
\vert\xi(\alpha,\theta^{(1)},\cdots,\theta^{(M)})\rangle=\sum_{i=1}^{M}\alpha_{i}\vert\psi_{i}(\theta^{(i)})\rangle,\label{eq:Ansatz_IQAE_General}
\end{equation}
where $\alpha_{i}\in\mathbb{C}$ and $\vert\psi_{i}(\theta^{(i)})\rangle$
are parametric quantum circuits. The aforementioned ansatz has been
employed in several works~\cite{mcclean2017hybrid,kyriienko2020quantum,parrish2019quantum,bespalova2020hamiltonian,huggins2020non,takeshita2020increasing,stair2020multireference,motta2020determining,seki2020quantum} and hence our algorithm shares features
with a few existing NISQ algorithms for finding the approximation
to ground state for a given Hamiltonian. Since our algorithm requires
calculation of matrix overlap elements, the relevant work work for
comparison are those based on computation of matrix overlaps~\cite{kyriienko2020quantum,parrish2019quantum,bespalova2020hamiltonian,huggins2020non,takeshita2020increasing,stair2020multireference,motta2020determining,seki2020quantum}. Unlike
the works in ~\cite{kyriienko2020quantum,parrish2019quantum,bespalova2020hamiltonian,huggins2020non,takeshita2020increasing,stair2020multireference,motta2020determining,seki2020quantum}, our approach does not require controlled multiqubit unitaries
for complicated tests such as the Hadamard test.

In particular, our algorithm shares some of the features with quantum
subspace expansion (QSE)~\cite{mcclean2017hybrid}, However, there are several key differences
between IQAE and QSE, which we enumerate here.
\begin{enumerate}
    \item While QSE works for fermionic excitation or Pauli unitaries, IQAE works for any k-local unitary as well as Pauli unitaries.
    \item The final step of IQAE is a quadratically constrained quadratic program, which admits a semidefinite relaxation.
    \item VQE is a special case of IQAE (see section \ref{sec: QAE_VQE} in SM for details).
    \item The Ansatz in IQAE is justified using imaginary time evolution and has connections with Krylov subspace. 
    \item IQAE is able to calculate the ground state for different Hamiltonian without requiring additional measurements of the quantum computer. This is achieved by varying the $\beta$ parameter that parameterizes the Hamiltonian during post-processing.
    \item IQAE can solve systems of thousands of qubits as demonstrated in the main text.
\end{enumerate}}

\section{Krylov subspace based algorithm} \label{sec: Krylov}
\revA{For a particular positive integral value of $K$, we have the following
Ansatz,
\begin{equation}
\vert\varsigma\left(\alpha\right)\rangle^{\left(K\right)}=\alpha_{0}\vert\psi\rangle+\alpha_{1}H\vert\psi\rangle+\cdots+\alpha_{K}H^{K}\vert\psi\rangle,\label{eq:Ansatz_Krylov_Ham}
\end{equation}
where $\alpha_{i}\in\mathbb{C}$ for $i\in\left\{ 0,1,\ldots,K\right\} .$
For the aforementioned Ansatz, the Hamiltonian ground state problem
reduces to the following optimization program,
\begin{equation*}
\text{minimize }  \boldsymbol{\alpha}^{T}\mathcal{D}_{H}^{(K)}\boldsymbol{\alpha}
\end{equation*}
\begin{equation}
\text{ subject to }  \boldsymbol{\alpha}^{T}\mathcal{E}_{H}^{(K)}\boldsymbol{\alpha}=1.
\label{eq:optimization_QCQP_Ham}
\end{equation}
For a particular value of $K$, the overlap matrices $\mathcal{D}_{H}^{(K)}$
and $\mathcal{E}_{H}^{(K)}$ are constructed using the Ansatz in equation
\ref{eq:Ansatz_Krylov_Ham} similar to the regular IQAE algorithm.
The matrix size scales linearly with $K$. However, the downside is
that calculating the overlap matrices would require complicated multi
qubit unitaries, which is not suitable for the devices available in
the next few years.}
\section{Justification for the Ansatz} \label{sec: justify}
For the sake of completeness, we reiterate the relevant content from the main text. Suppose the initial state ($0$-moment
state) $\vert\psi\rangle$ admits following representation in the
eigenbasis of the Hamiltonian $H$

\begin{equation}
\vert\psi\rangle=\sum_{i=1}^{\mathcal{N}}a_{i}\vert\phi_{i}\rangle\, ,\label{eq:initial_state_eigen_decomposition_ap}
\end{equation}
where $a_{i}\in\mathbb{C}$ for $i\in\{1,2,\cdots,\mathcal{N}\}$. Starting
with $\vert\psi\rangle$, if one applies $\exp\left(-\tau H\right)$
for some $\tau\geq0$, the normalized state is given by
\begin{equation}
\vert\gamma\rangle=\frac{e^{-\tau H}\vert\psi\rangle}{\langle\psi\vert e^{-2\tau H}\vert\psi\rangle}.\label{eq:ITE_state_1_ap}
\end{equation}
Using $e^{-\tau H}=\sum_{p=0}^{\infty}\frac{\left(-\tau H\right)^{p}}{p!}$,
we get
\begin{equation}
\vert\gamma\rangle=\frac{\sum_{p=0}^{\infty}\frac{\left(-\tau H\right)^{p}}{p!}\vert\psi\rangle}{\langle\psi\vert\sum_{p=0}^{\infty}\frac{\left(-2\tau H\right)^{p}}{p!}\vert\psi\rangle}.\label{eq:ITE_state_expanded_ap}
\end{equation}
In the asymptotic limit $\tau\rightarrow\infty$, $\vert\gamma\rangle\rightarrow\vert\phi_{1}\rangle.$
Let us define the operator 
\begin{equation}
\mathcal{O}^{K}\equiv\sum_{p=0}^{K}\frac{\left(-\tau H\right)^{p}}{p!},\label{eq:K-Approx_Operator-1_ap}
\end{equation}
for $K\geq0.$ Notice that $\mathcal{O}^{K}$ corresponds to the sum
of first $K$ terms of $e^{-\tau H}.$ Using $\mathcal{O}^{K},$ we
proceed to define
\begin{equation}
\vert\gamma_{K}\rangle\equiv\frac{\sum_{p=0}^{K}\frac{\left(-\tau H\right)^{p}}{p!}\vert\psi\rangle}{\langle\psi\vert \left(\sum_{p=0}^{K}\frac{\left(-\tau H\right)^{p}}{p!}\right)^2\vert\psi\rangle}.\label{eq:Evolved_state_ITE_K-Approx-1_ap}
\end{equation}
For $K\rightarrow\infty$, $\vert\gamma_{K}\rangle\rightarrow\vert\gamma\rangle.$
Using the expression for Hamiltonian as linear combination of unitaries,
it is easy to see that $\vert\gamma_{K}\rangle$ can be written as
linear combination of cumulative $K$-moment states, i.e,
\[
\vert\gamma_{K}\rangle=\sum_{\vert \chi_{i}\rangle\in\mathbb{CS}_{K}}\alpha_{i}\vert \chi_{i}\rangle
\]
where the combination coefficients $\alpha_{i}\in\mathbb{C}.$ The
aforementioned arguments justify the choice of Ansatz as a linear combination of cumulative $K$-moment states.

\noindent 
The probability of $\vert\gamma\rangle$ being ground state is 

\begin{equation}
P_{g}^{\gamma}\equiv\left|\langle\phi_{1}\vert\gamma\rangle\right|^{2}.\label{eq:prob_ground_gamma}
\end{equation}
Hence, we proceed to calculate

\[
\langle\phi_{1}\vert\gamma\rangle=\frac{\langle\phi_{1}\vert\sum_{p=0}^{\infty}\frac{\left(-\tau H\right)^{p}}{p!}\vert\psi\rangle}{\langle\psi\vert\sum_{p=0}^{\infty}\frac{\left(-2\tau H\right)^{p}}{p!}\vert\psi\rangle}
\]

\begin{equation}
=\frac{a_{1}\sum_{p=0}^{\infty}\frac{\left(-\lambda_{1}\tau\right)^{p}}{p!}}{\sum_{p=0}^{\infty}\frac{\left(-2\tau\right)^{p}}{p!}\langle\psi\vert H^{p}\vert\psi\rangle}.\label{eq:overlap_ground_amp_1}
\end{equation}
Defining $E^{P}$ to be the expectation value of $H^{p}$ with respect
to $\vert\psi\rangle$ i.e 
\begin{equation}
E^{p}\equiv\langle\psi\vert H^{p}\vert\psi\rangle,\label{eq:Expectation_Moments}
\end{equation}
we get,

\begin{equation}
\langle\phi_{1}\vert\gamma\rangle=\frac{a_{1}\sum_{p=0}^{\infty}\frac{\left(-\lambda_{1}\tau\right)^{p}}{p!}}{\sum_{p=0}^{\infty}\frac{\left(-2\tau\right)^{p}}{p!}E^{p}}.\label{eq:amp_ground_exp_moments}
\end{equation}
Using \ref{eq:amp_ground_exp_moments}, we get

\begin{equation}
P_{g}^{\gamma}=\vert a_{1}\vert^{2}\left|\frac{\sum_{p=0}^{\infty}\frac{\left(-\lambda_{1}\tau\right)^{p}}{p!}}{\sum_{p=0}^{\infty}\frac{\left(-2\tau\right)^{p}}{p!}E^{p}}\right|^{2}.\label{eq:prob_computed_all}
\end{equation}
Let us define

\begin{equation}
A_{K}\equiv\frac{\sum_{p=0}^{K}\frac{\left(-\lambda_{1}\tau\right)^{p}}{p!}}{\sum_{p_1=0}^{K}{\sum_{p_2=0}^{K}\frac{\left(-\tau\right)^{p_1+p_2}}{p_1!p_2!}E^{p_1+p_2}}}.\label{eq:Coefficients_Prob}
\end{equation}
 In the asymptotic limit $K\rightarrow\infty,$ $\vert a_{1}\vert^{2}\vert A_{K}\vert^{2}\rightarrow P_{g}^{\gamma}.$
We define the error term in the finite case as

\begin{equation}
\epsilon_{K}\equiv\left|P_{g}^{\gamma}-\vert a_{1}\vert^{2}\vert A_{K}\vert^{2}\right|.\label{eq:error_term}
\end{equation}

The error term $\epsilon_{K}$ depends on the relative importance of the consecutive terms, i.e., $\frac{A_K}{A_{K+1}}.$ In future, one can analyze $\frac{A_K}{A_{K+1}}$ to provide theoretical guarantees on the accuracy of the IQAE algorithm depending on the iteration number $K$. 

\noindent 
It is natural to assume that the initial state $\vert\psi\rangle\in\mathbb{C}^{\mathcal{N}}$
belongs to the set of pure states equipped with Haar-induced measure.
We refer to such a $\vert\psi\rangle$ as a random quantum state.
Given a Hamiltonian $H,$ the closed form expression for $\langle\psi\vert H^{p}\vert\psi\rangle$
is given by \cite{venuti2013probability}

\begin{equation}
E^{p}=\frac{\left(\mathcal{N}-1\right)!}{\left(\mathcal{N}+p-1\right)!}\sum_{\pi\in S_{p}}\text{tr}\left(\mathbb{P}_{\pi}H^{\otimes p}\right).\label{eq:moment_expectation_Haar}
\end{equation}
Here, $S_{p}$ denotes the symmetric group of $p$ elements and $\mathbb{P}_{\pi}$
enables the permutation operator $\pi$ in the corresponding Hilbert
space. Using the expression in \ref{eq:moment_expectation_Haar} for
$E^{p}$, for $p=1,2,3$, one gets

\[
E^{1}=\frac{\text{tr}\left(H\right)}{\mathcal{N}},
\]

\[
E^{2}=\frac{\text{tr}\left(H\right)^{2}+\text{tr}\left(H^{2}\right)}{\mathcal{N}\left(\mathcal{N}+1\right)},
\]

\[
E^{3}=\frac{\text{tr}\left(H\right)^{3}+3\text{\text{tr}}\left(H^{2}\right)\text{tr}\left(H\right)+2\text{ tr}\left(H^{3}\right)}{\mathcal{N}\left(\mathcal{N}+1\right)\left(\mathcal{N}+2\right)}.
\]
If all the eigenvalues of $H$ are non-degenerate, one obtains the
following expression for $\langle\psi\vert H^{p}\vert\psi\rangle$,

\begin{equation}
E^{p}=\frac{1}{{N+p-1 \choose p}}\sum_{j=1}^{N}\frac{\left(\lambda_{j}\right)^{N+p-1}}{\prod_{j\neq k}\left(\lambda_{j}-\lambda_{k}\right)}.\label{eq:Moments_Haar_non_degenerate}
\end{equation}

\section{Numerical experiments} \label{sec: Numerical}
In this section, we perform numerical simulations for various example problems of finding ground states with IQAE.

\begin{figure}
	\centering
	\subfigimg[width=0.22\textwidth]{}{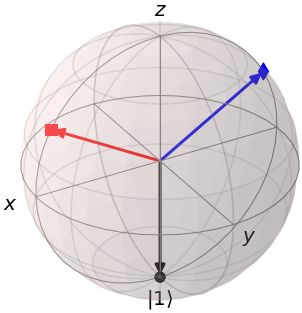}
	\caption{Illustration of IQAE for finding ground state of a single qubit Hamiltonian.
	}
	\label{Bloch}
\end{figure}

First, we solve a simple one-qubit Hamiltonian problem $H_\text{B}=\sigma^z$ in Fig.\ref{Bloch}. 
In step $1$ we choose an arbitrary initial state (0-moment) $\mathbb{S}_0=\{\ket{\chi_0}\}$ (red vector). Then, using the set of unitaries that make up the Hamiltonian $\{\sigma^z\}$, we generate the $1$-moment states  $\mathbb{S}_1=\{\ket{\chi_1}\}$ with $\ket{\chi_1}=\sigma^z\ket{\chi_0}$ (blue vector). The union of the two moments give us the cumulative $1$-moment states $\mathbb{CS}_1=\{\ket{\chi_0},\ket{\chi_1}\}$. In step 2, we calculate the overlap matrices $\mathcal{D}_{n,m}^{(1)}=\bra{\chi_n}H_\text{B}\ket{\chi_m}$, $\mathcal{E}_{n,m}^{(1)}=\braket{\chi_n}{\chi_m}$ (Eq.\ref{eq:D_matrix},\ref{eq:E_matrix}) using the quantum computer. 
Then, in step $3$, we calculate the coefficients $\boldsymbol{\alpha}$ to compose the ground state $\ket{\phi_1}=\alpha_0\ket{\chi_0}+\alpha_1\ket{\chi_1}$ (black vector) using program \eqref{eq:P1}. Here, IQAE converges to the ground state after one iteration as the Krylov subspaces closes for order $K=1$ due to $\mathbb{S}_2=\{\sigma^z\sigma^z\ket{\chi_0}\}=\{\ket{\chi_0}\}=\mathbb{S}_0$.

\begin{figure}[htbp]
	\centering
	\subfigimg[width=0.24\textwidth]{a}{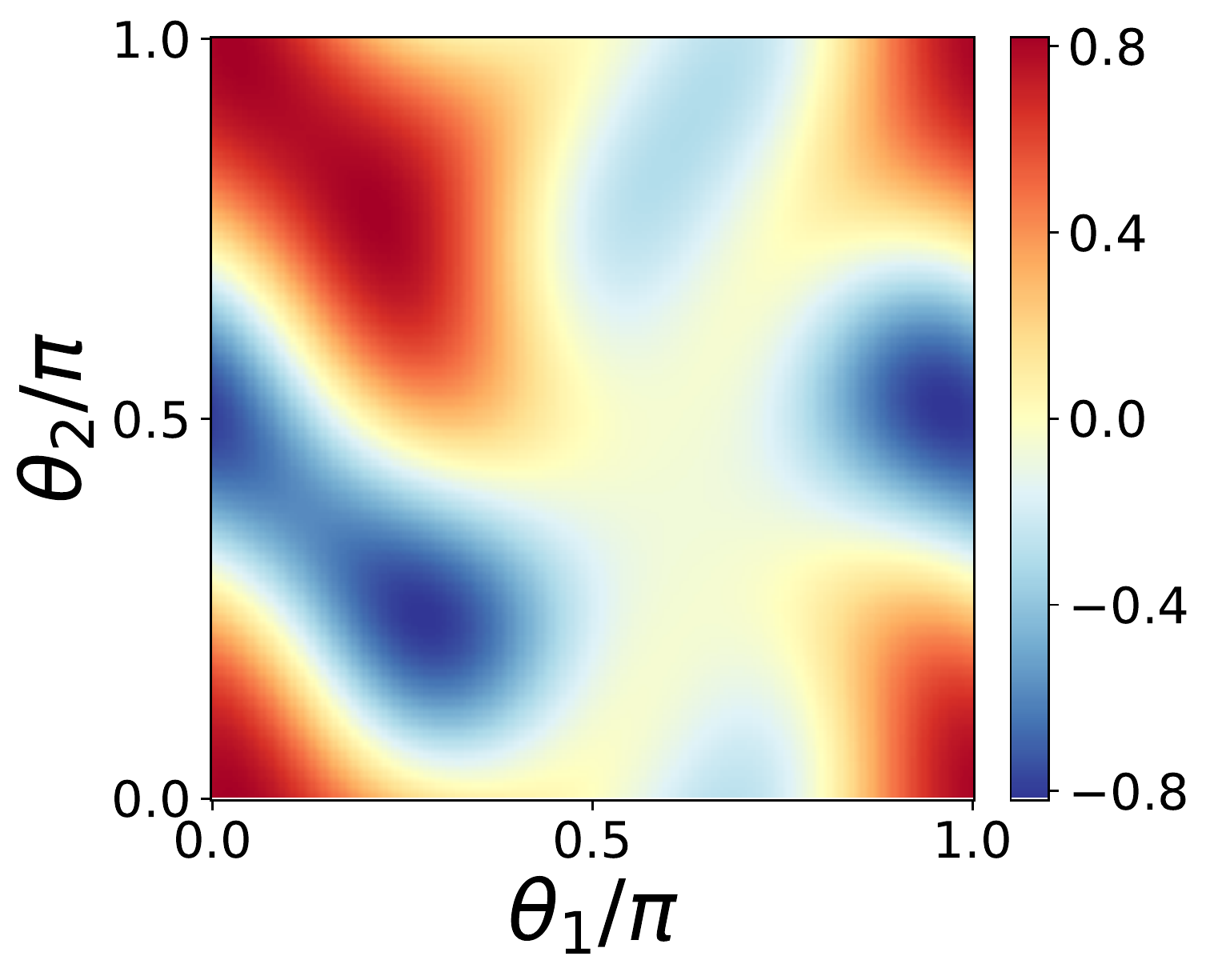}\hfill
	\subfigimg[width=0.24\textwidth]{b}{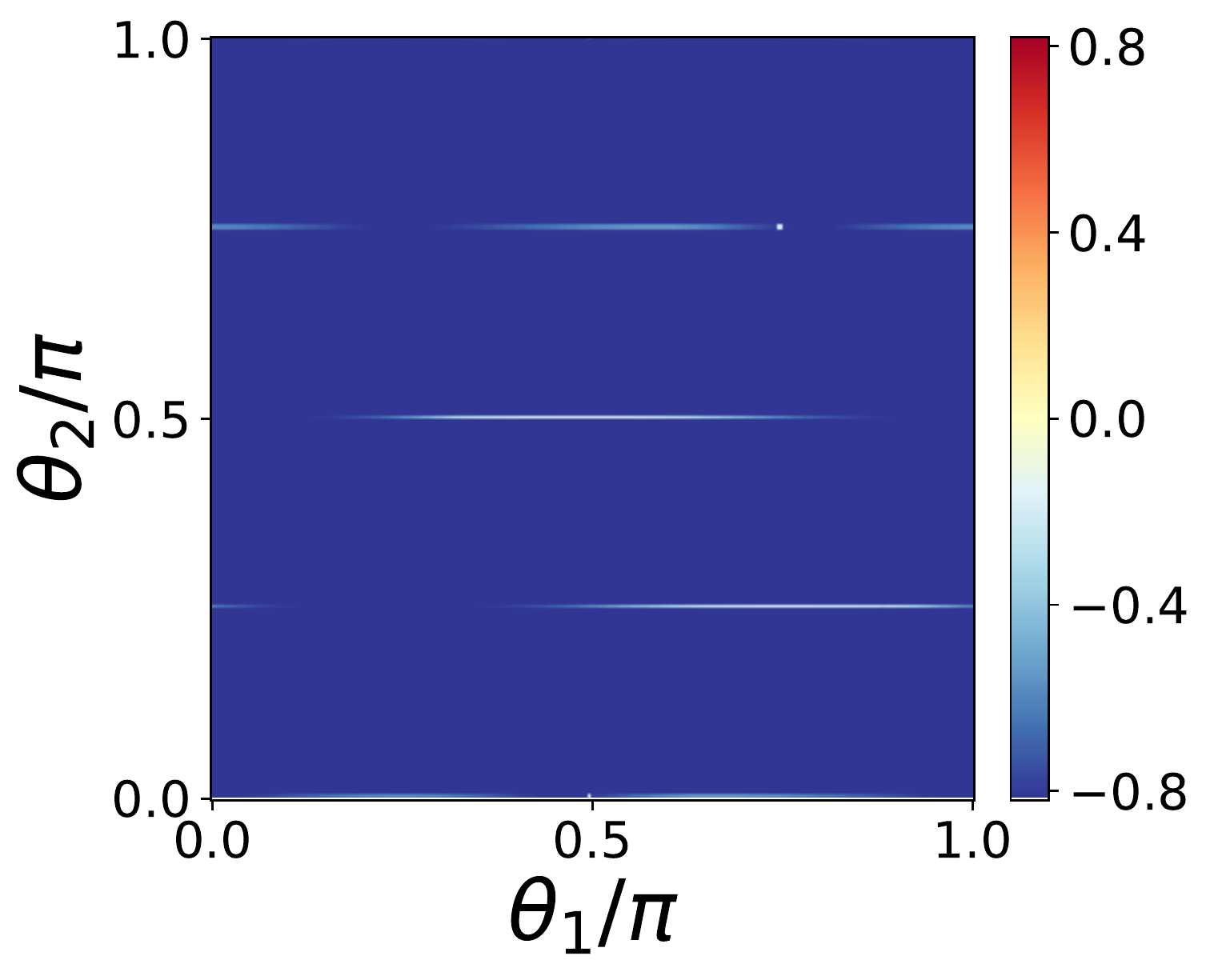}
	\caption{IQAE for H2 molecule with hardware efficient Ansatz parameterized by angles $\theta_1$ and $\theta_2$. \idg{a} Energy landscape $\bra{0}U^\dagger(\theta_1,\theta_2)H_2 U(\theta_1,\theta_2)\ket{0}$. \idg{b} Optimised energy found by IQAE for given initial state $U(\theta_1,\theta_2)\ket{0}$, with exact ground state energy $E_\text{g}=-0.8246$ found for nearly all states.}
	\label{H2}
\end{figure}

We apply IQAE to a standard problem of quantum chemistry, the $H_2$ molecule, in Fig.\ref{H2}. We can approximate the Hamiltonian as $H_2=\beta_1\sigma_1^z+\beta_2\sigma_2^z+\beta_3\sigma_1^x\sigma_2^x$ \cite{bravyi2017tapering,yamamoto2019natural} for $\beta_1=\beta_2 = 0.4$ and $\beta_3 = 0.2,$ and choose a hardware efficient variational Ansatz $U(\theta_1,\theta_2)=\text{R}^y_1(\theta_2)\text{R}^y_2(\theta_2)\text{CNOT}(1,2)\text{R}^y_1(\theta_1)\text{R}^y_2(\theta_1)\ket{00}$, where $\text{R}^y_n(\theta)=\exp({-i\theta/2\sigma_n^y})$ and $\text{CNOT}(c,t)$ is the CNOT gate with control qubit $c$ and target $t$, giving us the variational wavefunction $\ket{\psi}=U(\theta_1,\theta_2)\ket{0}$.
In Fig.\ref{H2}a we show the energy landscape $\bra{\psi}H_2\ket{\psi}$, which features a non-convex landscape with local minimas of the cost function.
We now apply IQAE and determine the cumulative $1$-moment states $\mathbb{CS}_1=\{\ket{\psi},\sigma_1^z\ket{\psi},\sigma_2^z\ket{\psi},\sigma_1^x\sigma_2^x\ket{\psi}\}$.
Then we  calculate the $\mathcal{D}^{(1)}$ and $\mathcal{E}^{(1)}$ matrices and find the optimised $\boldsymbol{\alpha}$ with program \eqref{eq:P1}.
We repeat this procedure for every initial Ansatz state $\ket{\psi(\theta_1,\theta_2)}$ parameterized by $\theta_1,\theta_2$. We show the resulting optimized energy found by IQAE in Fig.\ref{H2}b.
For nearly every initial state, we can find the optimal ground state. We find stripes of higher energy only for very particular choices of $\theta_2$, where the initial state has no overlap with the ground state and the moment expansion is unable to reach the ground state.

Now, we study a anti-ferromagnetic spin system, the ring of disagrees, which has been studied intensively for QAOA \cite{farhi2014quantum}. Here, the Hamiltonian is
\begin{equation}
H_\text{ROD}=\frac{1}{2}\sum_{i=1}^N(1-\sigma^z_i\sigma^z_{i+1})\,
\end{equation}
where $\sigma^\alpha_i$, $\alpha\in \{x,y,z\}$ is the Pauli operator  acting on qubit $i$. This Hamiltonian contains only $\sigma^z$ operators and thus is diagonal in the computational basis.

Further, we study a hallmark quantum spin model, the transverse Ising model
\begin{equation}
H_\text{ising}=\frac{J}{2}\sum_{i=1}^N\sigma^x_i\sigma^x_{i+1}-\frac{1}{2}\sum_{i=1}^N\sigma^z_i
\end{equation}
which describes a spin system with an applied magnetic field.
We apply IQAE to find the ground state of these Hamiltonians. We note that the Hamiltonians we are considering are sums of Pauli strings, and thus the measurements for the elements of the $\mathcal{D}$ and $\mathcal{E}$ matrices (Eq.\ref{eq:D_matrix},\ref{eq:E_matrix}) involve only  measurements of Pauli operators, which can be easily performed on existing NISQ hardware.
We investigate three types of Ans\"atze. First, the Ansatz from \cite{mcclean2018barren} (see also main text) $\ket{\psi(\boldsymbol{\theta})}$, parameterized by a set of angles $\boldsymbol{\theta}$. For many layers $d$ and random angles $\boldsymbol{\theta}$, this type of Ansatz produces a randomized state with vanishing mean as well as variance of the gradients $\text{var}(\partial_{\boldsymbol{\theta}}\bra{\psi(\boldsymbol{\theta})}H\ket{\psi(\boldsymbol{\theta})})$.
Next, an initial Ansatz state that is a simple product state $\ket{+}^{\otimes N}$, where $\ket{+}$ indicates the eigenstate with positive eigenvalue of the $\sigma^x$ operator. Finally, a QAOA Ansatz of $p$ layers $\prod_{k=1}^p\exp(-i\beta_k\sum_j\sigma^x_j)\exp(-i\gamma_kH)\ket{+}^{\otimes N}$, where  $\gamma,\beta$ parameters have been optimized to minimize the problem Hamiltonian by using the COBYLA solver. For $p\rightarrow \infty$, this Ansatz converges to the ground state. We choose $p=1$, such that the Ansatz is an approximation to the actual target ground state.

\begin{figure}[htbp]
	\centering
	\subfigimg[width=0.24\textwidth]{a}{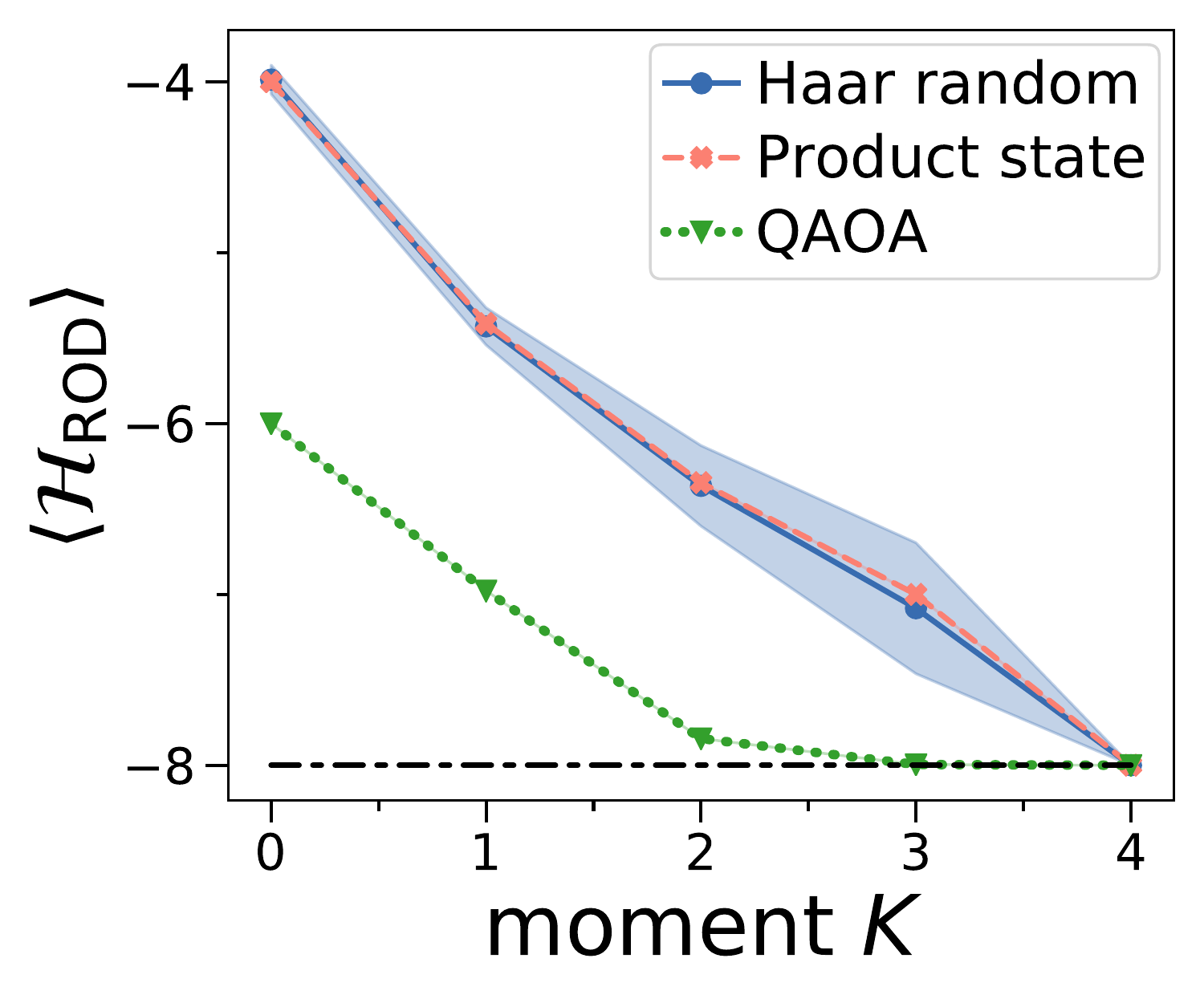}\hfill
	\subfigimg[width=0.24\textwidth]{b}{energyresQAEN8D1C2o2r1c2S0o3000N100t1t2t3g0h-1_0.pdf}
	\caption{IQAE cost function plotted against order of moment expansion for two different problem Hamiltonians \idg{a} Ring of disagree \idg{b} Transverse Ising model ($J=1$). Both systems have $N=8$ qubits. We show three Ans\"atze: Haar random variational states that exhibit barren plateaus with 200 layers that are averaged over 100 instances, a simple product state as well as QAOA Ansatz of depth $p=1$ with pre-optimized parameters. Shaded lines indicate standard deviation of the sampled cost function.}
	\label{Expansion}
\end{figure}

We run the IQAE iterations up to a specified order $K$: In step $1$, we expand the initial wavefunction with the cumulative $K$-moment states $\mathbb{CS}_K$. Then, in step $2$, we calculate the $\mathcal{D}^{(K)}, \mathcal{E}^{(K)}$ matrices. In step $3$, the optimised $\boldsymbol{\alpha}$ for the superposition state $\vert\xi\left(\boldsymbol{\alpha}\right)\rangle^{\left(1\right)}$ are found, and the expectation value of the energy $\langle H\rangle$ is calculated. The energy for moment $K$ and different Ans\"atze is shown for ring of disagrees in Fig.\ref{Expansion}a, and transverse Ising in Fig.\ref{Expansion}b.
We find that the QAOA Ansatz works best in both cases.
The moment $K$ at which a specific Ansatz converges depends on the Ansatz. Curiously, we find that for transverse Ising, the Haar random state converges at $K=3$, while the product state with lower initial energy only converges at $K=4$. For ring of disagrees, we find no difference between random and product Ansatz.

\section{IQAE and imaginary time evolution}\label{sec:imaginary}
\begin{figure}[htbp]
	\centering
	\subfigimg[width=0.24\textwidth]{}{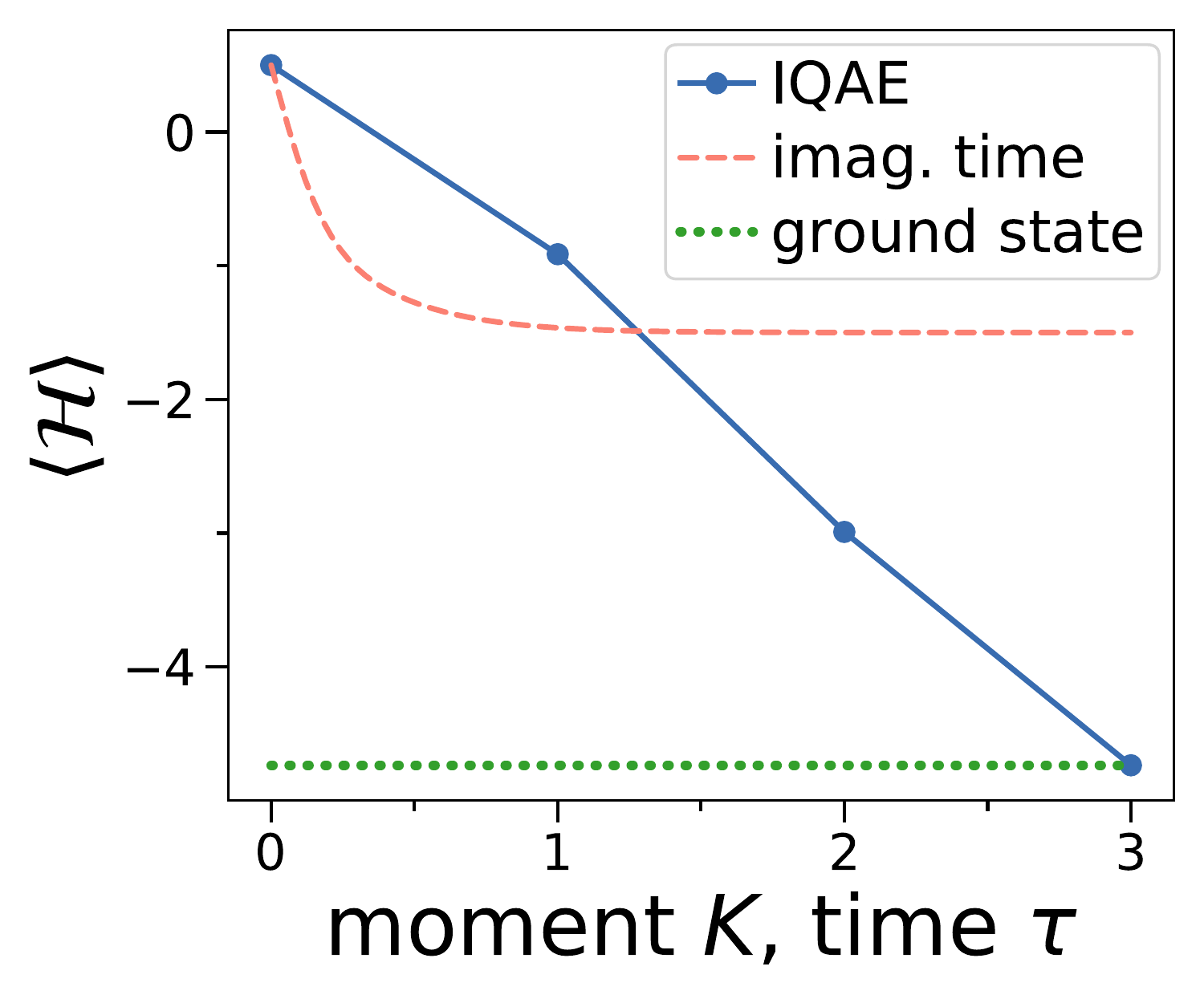}\hfill
	\caption{Heisenberg XXZ model for IQAE and imaginary time evolution for varying moments $K$ or time evolution $\tau$. The Ansatz state has no overlap with target ground state.  We choose an initial Fock state with number of spin excitations $\mathcal{M}=1$, $N=6$ and $\Delta=0.5$.}
	\label{XXZ}
\end{figure}

We look at the Heisenberg XXZ model 
\begin{equation}
H_\text{XXZ}=\frac{1}{2}\sum_{i=1}^N(\sigma^x_i\sigma^x_{i+1}+\sigma^y_i\sigma^y_{i+1} + \Delta\sigma^z_i\sigma^z_{i+1})\,\,
\end{equation}
where $\Delta$ controls the spin interaction strength  between neighboring qubits. In this model, the number of spin excitations $\mathcal{M}=\sum_i\sigma_z$ commutes with the Hamiltonian $[H_\text{XXZ},\mathcal{M}]=0$ and is a conserved quantity. 
In the main text, we referenced the method of imaginary time evolution $\ket{\psi(\tau)}=\exp(-\tau H)\ket{\psi(0)}=\sum_{i=1}^\mathcal{N}\exp(-\tau \lambda_i)\braket{\phi_i}{\psi(0)}\ket{\phi_i}$, for a given initial state $\ket{\psi(0)}$, the eigenstate $\ket{\phi_i}$ and eigenvalue $\lambda_i$ of the Hamiltonian for which we want to find the groundstate. This non-unitary evolution leads to exponential decay of all states, with the ground state decaying the slowest. After sufficient evolution time $\tau$, only the lowest energy eigenstate remains. However, if the ground state $\ket{\phi_1}$ has no overlap with initial state $\braket{\phi_1}{\psi(0)}=0$ (e.g. because the initial state is in a different symmetry sector), it is not possible to find the correct ground state.
\revA{To demonstrate this problem in an instructive way,} we prepare an initial state which is an eigenstate of $\mathcal{M}$ and the conserved quantity does not match the ground state $\bra{\psi(0)}\mathcal{M}\ket{\psi(0)}\ne\bra{\phi_1}\mathcal{M}\ket{\phi_1}$. Thus, $\braket{\phi_1}{\psi(0)}=0$.  Here, we compare imaginary time evolution and IQAE (see Fig.\ref{XXZ}). We find that while imaginary time evolution does not converge, IQAE converges to the ground state of the system. 
This can be explained as follows: The IQAE moments are constructed from a product of the Hamiltonian unitaries. While the total Hamiltonian conserves the $\mathcal{M}$, the individual unitaries may not do so. Thus, the IQAE moments can consist of terms breaking the symmetry, allowing for convergence.
\revA{While the ansatz with the right symmetry could be easily constructed in the Heisenberg model, this can be difficult in more complex Hamiltonians with non-trivial symmetries. Here, IQAE can provide a path to the correct ground state even when imaginary time evolution fails.}
The found energy via imaginary time evolution and IQAE is shown in Fig.\ref{XXZ}.

\section{IQAE subsumes the Variational Quantum Eigensolver} \label{sec: QAE_VQE}
In this section, we discuss how IQAE can subsume the Variational Quantum Eigensolver (VQE).
In VQE, for a given Ansatz state $\ket{\psi(\boldsymbol{\theta})}=U(\boldsymbol{\theta})\ket{0}$ the angles $\boldsymbol{\theta}$ that parameterize the Ansatz are adjusted to optimize a given Hamiltonian $H$. A common approach in VQE is to find the optimal $\boldsymbol{\theta}^*$ by using the gradient of the energy $\partial_{\theta}\bra{\psi(\theta)}H\ket{\psi(\theta)}$.
IQAE uses an alternative approach: $\boldsymbol{\theta}$ remains fixed, instead the initial state $\ket{\psi(\boldsymbol{\theta})}$ is expanded into moments from the cumulative $K$-moment states $\mathbb{CS}_{K}$, with $\vert\xi\left(\boldsymbol{\alpha}\right)\rangle^{\left(K\right)}=\sum_{\vert \chi_{i}\rangle\in\mathbb{CS}_{K}}\alpha_{i}\vert \chi_{i}\rangle$ with $\boldsymbol{\alpha}$ chosen such that it minimized $H$.
We now combine both IQAE and VQE together. 
First, we use the IQAE up to a given moment $K$ and calculate the matrix elements for the expanded state $\mathcal{D}_{n,m}(\boldsymbol{\theta})$, $\mathcal{E}_{n,m}(\boldsymbol{\theta})$, where we dropped the superscript $(K)$ for convenience. If the Hamiltonian $H=\sum_i \beta_i U_i$ consists of $U_i$ which are Pauli strings, then $\mathcal{D}_{n,m}$, $\mathcal{E}_{n,m}$ are measurements of Pauli operators on the variational state $\ket{\psi(\boldsymbol{\theta})}$. We then find $\boldsymbol{\alpha}$ such that minimizes the energy. 
Then, we use the concept of VQE and calculate the gradient of the energy of the superposition state
\begin{align*}
&\partial_\theta \bra{\xi}H\ket{\xi}=\\
&\sum_{n,m}\alpha_n^*\alpha_m(\partial_\theta \mathcal{D}_{n,m})+(\partial_\theta\alpha_n^*)\alpha_m \mathcal{D}_{n,m}+\alpha_n^*(\partial_\theta\alpha_m) \mathcal{D}_{n,m}\,.
\end{align*}
The gradients $\partial_\theta \mathcal{D}_{n,m}$ and $\partial_\theta \mathcal{E}_{n,m}$ can be measured directly with the quantum computer by e.g. the shift rule~\cite{mitarai2018quantum}. We now derive the gradient of $\boldsymbol{\alpha}$. With the condition $1=\sum_{n,m}\alpha_n^*\alpha_m \mathcal{E}_{n,m}$ we find
$0=\sum_{n,m}(\partial_\theta \alpha_n^*)\alpha_m \mathcal{E}_{n,m}+ \alpha_n^*(\partial_\theta\alpha_m) E_{n,m}+\alpha_n^*\alpha_m \partial_\theta \mathcal{E}_{n,m}$. By re-ordering we get
$0=\sum_{n}\alpha_n^*(\sum_m\partial_\theta \alpha_m \mathcal{E}_{n,m}+\frac{1}{2}\alpha_m\partial_\theta \mathcal{E}_{n,m})+\text{h.c}$, where $\text{h.c}$ indicates the hermitian conjugate. This equation is in general only fulfilled if the sum inside the bracket is zero. This condition rewritten into matrix form gives $\mathcal{E}(\partial_\theta \boldsymbol{\alpha}) +\frac{1}{2}(\partial_\theta\mathcal{E}) \boldsymbol{\alpha}=0$. This equation is solved by $\partial_\theta \boldsymbol{\alpha}=-\frac{1}{2}\mathcal{E}^+(\partial_\theta \mathcal{E})\boldsymbol{\alpha}$, where $\mathcal{E}^+$ is the pseudo inverse of $\mathcal{E}$. The gradient can then be used to update the Ansatz wavefunction via any gradient based optimization method.

\section{Linear Combination of Unitaries as a Ring} \label{sec: Ring}
It is important to notice that the set of linear combination of unitaries
forms a ring. For the sake of completion, we provide the mathematical
definition of Ring here.
\begin{defn} A set $S$ with two binary operations $"+"$ and $"*"$ is
a ring if it satisfies the following conditions.
\begin{enumerate}
\item $\left(a_{1}+a_{2}\right)+a_{3}=a_{1}+\left(a_{2}+a_{3}\right)$ $\forall a_{1},a_{2},a_{3}\in S$.
\item $a_{1}+a_{2}=a_{2}+a_{1}$ $\forall a_{1},a_{2}\in S.$
\item There exists an element $0\in S$ such that $a+0=a$ $\forall a\in S.$
\item For every $a\in S$, there exists $-a\in S$ such that $a+\left(-a\right)=0.$
\item $\left(a_{1}*a_{2}\right)*a_{3}=a_{1}*\left(a_{2}*a_{3}\right)$ $\forall a_{1},a_{2},a_{3}\in S.$
\item There exists an element $1\in S$ such that $a*1=a$ $\forall a\in S.$
\item $a_{1}*\left(a_{2}+a_{3}\right)=\left(a_{1}*a_{2}\right)+\left(a_{1}*a_{3}\right)$
$\forall a_{1},a_{2},a_{3}\in S.$
\item $\left(a_{2}+a_{3}\right)*a_{1}=\left(a_{2}*a_{1}\right)+\left(a_{3}*a_{1}\right)$
$\forall a_{1},a_{2},a_{3}\in S.$
\end{enumerate}
\end{defn}
The set of linear combination of unitaries can be formally described
as $\mathbb{LCU}=\left\{ \sum_{i=1}^{r}\alpha_{i}U_{i}\vert\alpha_{i}\in\mathbb{C},r\in\mathbb{N}\text{  and }U_{i}\in SU(\mathcal{N})\right\} $, where $\mathbb{N}$ denotes the set of natural numbers. It is straightforward
to check that $\mathbb{LCU}$ satisfies all of the required conditions
for a ring.
\section{Measurements} \label{sec: measurements}
For the purposes of this section, a unitary $U$ will be referred $k$-local if it acts non-trivially on at most $k$ qubits. We assume $k=\mathcal{O}(poly(log(N))$. The step $2$ of iteration $K$ for some positive integer $K$, the IQAE algorithm requires computation of matrix elements
of the form $\langle\psi^\vert U\vert\psi\rangle,$
where $U$ is a k-local unitary matrix (since $U$ is product of at most $2K+1$
k-local unitary matrices). By invoking the following result from Mitarai \textit{et.
al.} \citep{mitarai2019methodology}, we guarantee an efficient
computation of the overlap matrices on a quantum computer without the use of the
Hadamard test. 
\begin{fact} \cite{mitarai2019methodology} \label{fact: hadamard}
Let k be an integer such that $k=\mathcal{O}(poly(log(N))$, where
$N$ is the number of qubits and $\ket{\psi}$ be an $N$-qubit quantum state. For any k-local quantum gate $U,$ it
is possible to estimate $\langle\psi \vert U\vert \psi \rangle$
up to precision $\epsilon$ in time $\mathcal{O}\left(\nicefrac{k^{2}2^{k}}{\epsilon^{2}}\right)$
without the use of the Hadamard test, with classical preprocessing
of time $\mathcal{O}(poly(logN)).$
\end{fact}
We proceed to discuss the methodology suggested in \citep{mitarai2019methodology}
to calculate the required matrix elements. For detailed analysis,
refer to \citep{mitarai2019methodology}. Since $U$ is a $k$-local
unitary, it can be decomposed as $U=\otimes_{q=1}^{Q}U_{q}$ such
that $U_{q}$ acts on $k_{q}$ qubits. Clearly, $U_{q}$ is a $2^{k_{q}}\times2^{k_{q}}$
matrix. Suppose the eigenvalues of $U_{q}$ are $\left\{ \exp\left(i\phi_{q,m}\right)\right\} _{m=0}^{2^{k_{q}}-1}.$
Using the integers $m_{q}=0,\cdots,2^{k_{q}}-1,$ let us denote the
computational basis of each subsystem by $\vert m_{q}\rangle.$ We
diagonalize each $U_{q}$ and obtain some unitary matrix $V_{q}$
such that $U_{q}=V_{q}^{\dagger}T_{q}V_{q},$ where $T_{q}=\sum_{m=0}^{2^{k_{q}}-1}\exp\left(i\phi_{q,m}\right).$
Since $k=\mathcal{O}(poly(log(N))$, the aforementioned diagonalized
can be performed in polynomial time. A simple calculation gives,
\begin{multline}
 \langle\psi\vert U\vert\psi\rangle=\sum_{m_{1}=0}^{2^{k_{1}}-1}\cdots\sum_{m_{Q}=0}^{2^{k_{Q}}-1}\left(\prod_{q=1}^{Q}\exp\left(i\phi_{q,m_{q}}\right)\right)\\
\times\left|\left(\otimes_{q=1}^{Q}\langle m_{q}\vert\right)\left(\otimes_{q=1}^{Q}V_{q}\right)\vert \psi \rangle\right|^{2}.\label{eq:matrix_elements_evaluation}
\end{multline}
Thus, one can evaluate  $\langle\psi^\vert U\vert\psi\rangle$
by calculating the probability of getting $\otimes_{q=1}^{Q}\vert m_{q}\rangle$
from the measurement of $\left(\otimes_{q=1}^{Q}V_{q}\right)\vert \psi \rangle$
in the computational basis.
\begin{figure}[htbp]
\includegraphics[scale=0.2]{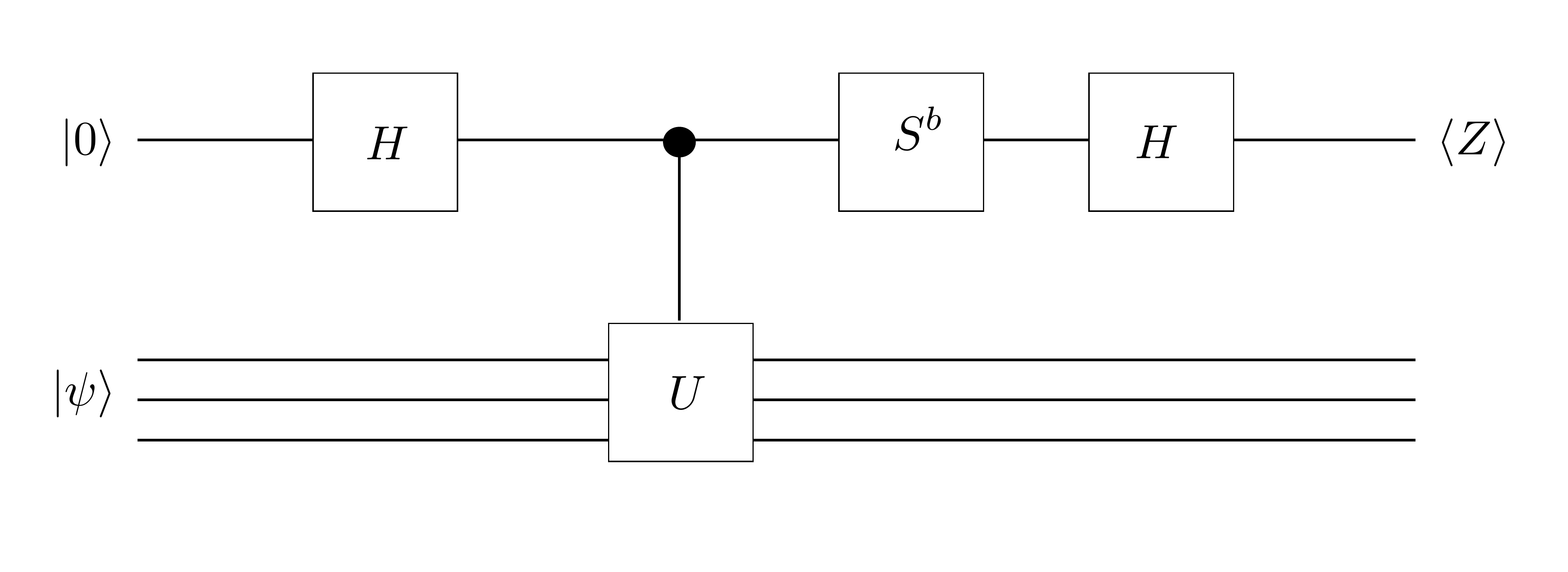}
\caption{The above figure shows a simple Hadamard test circuit for measuring
the real and imaginary part of $\langle \psi \vert U\vert \psi \rangle$
for any arbitrary $n$ qubit unitary $U$. The Hadamard gate has been
represented by $H$ and $S$ represents $e^{-\nicefrac{\iota\pi Z}{4}}.$
When $b=0,$ we get $\left\langle Z\right\rangle =\text{Re}\left\langle \psi \vert U\vert \psi \right\rangle .$
For $b=1,$ we get $\left\langle Z\right\rangle =\text{Im}\left\langle \psi\vert U\vert \psi \right\rangle .$
Since implementing controlled-unitaries is challenging in the NISQ
era, the use of Hadamard test as a subroutine is highly discouraged
while designing NISQ-friendly quantum algorithms. Using results from
\cite{mitarai2019methodology}, we guarantee an efficient computation of the overlap matrices required in the IQAE algorithm on a quantum computer without the use
of the Hadamard test.}
\label{fig: Hadamard}
\end{figure}

If the unitaries defining the Hamiltonian are Pauli strings over $N$ qubits, one can use the following Lemma \cite{huang2019near} to provide an estimate of the number of measurements needed to achieve a given desired accuracy. 
\begin{lem} \cite{huang2019near}
Let $\epsilon>0$ and $P_{q}$ be a  Pauli string over $N$
qubits. Let multiple copies of an arbitrary $N$-qubit quantum state
$\vert\psi\rangle$ be given. The expectation value $\langle\psi\vert P_{q}\vert\psi\rangle$
can be determined to additive accuracy $\epsilon$ with failure probability
at most $\delta$ using $\mathcal{O}\left(\frac{1}{\epsilon^{2}}\log\left(\frac{1}{\delta}\right)\right)$
copies of $\vert\psi\rangle.$
\end{lem}
We now discuss how to measure the matrix elements of $\mathcal{D}$ and $\mathcal{E}$ for the special case where the components $U_i$ of the Hamiltonian $H=\sum_i\beta_i U_i$ are Pauli strings $\bigotimes_{j=1}^N\gamma_j$, with $\gamma\in \{\mathbb{1},\sigma^x,\sigma^y,\sigma^z\}$. 
For a $K$-moment state for Ansatz $\ket{\psi}$, the elements to calculate are
\begin{align*}
\mathcal{D}_{n,m}&=\sum_i\beta_i\bra{\psi}U_{n_1}^\dagger\dots U_{n_K}^\dagger U_i U_{m_K}\dots U_{m_1}\ket{\psi}\\ \mathcal{E}_{n,m}&=\bra{\psi}U_{n_1}^\dagger\dots U_{n_K}^\dagger U_{m_K}\dots U_{m_1}\ket{\psi}. \numberthis\label{Eq:Pauli_matrix}
\end{align*}
Now, each overlap element is a product of a set of some Pauli strings $P_{q}$ to be measured on state $\ket{\psi}$, with $\bra{\psi}\prod_{q} P_{q}\ket{\psi}$. 
The product rule of Pauli operators states that $\sigma^i\sigma^j=\delta_{ij}\mathbb{1}+i\epsilon_{ijk}\sigma^k$, where $\sigma^1=\sigma^x,\, \sigma^2=\sigma^y,\, \sigma^3=\sigma^z$, $\delta_{ij}$ is the Kronecker delta and $\epsilon_{ijk}$ the Levi-Civita symbol. Thus, a product of two Pauli strings $P_q P_p=a P_s$ is again a Pauli string $P_s$, with a prefactor $a \in \{+1,-1,+i,-i\}$. 
To calculate the matrix elements on the quantum computer, first one has to evaluate which Pauli string corresponds to the product of unitaries in Eq.\ref{Eq:Pauli_matrix} and the corresponding prefactor. Then, the expectation value of the resulting Pauli string is measured for the Ansatz state $\bra{\psi}P_q\ket{\psi}$. This observable is a hermitian operator, and can be easily measured by rotating each qubit into the computational basis corresponding to the Pauli operator. Finally, the expectation value of the measurement is multiplied with the prefactor $a$.

\section{Future work} \label{sec: future}
In future studies, it would be very interesting to find ways to systematically reduce the number
of quantum states defining the Ansatz via methods such as a Ansatz tree structure~\cite{huang2019near}.
IQAE is well suited for problems that feature a Krylov subspace that closes quickly, i.e. $H^{K+1}\ket{\psi}\in span\{\ket{\psi},H\ket{\psi},\dots,H^K\ket{\psi}\}$ for $K$ small.
As an example, we simulate the ground state of a Hamiltonian consisting of multi-body Pauli strings for thousands of qubits. To facilitate classical simulation, we restricted as ansatz a product state. This problem could be run with an entangled ansatz beyond classical simulability to showcase the power of quantum computers and IQAE. We note that for other NISQ algorithms such as VQE it would be challenging to represent and optimize so many qubits and multi-body terms.
Another important example in many-body physics are quantum many-body scars. These quantum many-body scars often feature Krylov subspaces that close quickly and could be simulated with our algorithm~\cite{serbyn2021quantum}.
It is important to note that the set of linear combination of unitaries
forms a ring (see section \ref{sec: Ring} in SM) \revA{and the Krylov subspace idea is related to minimal polynomial of a matrix. Thus,} an in-depth mathematical analysis
could deliver intriguing insights towards systematically reducing
the number of quantum states defining the Ansatz. \revA{We believe that the ring structure might help understand the set of the linear combination of unitaries in terms of the generators of the ring.} We speculate that
techniques similar to Feynman diagrams could be invented to find
out a subset of highly contributing quantum states within the space of ansatz states. In future, it would be fascinating
to study the implications of our algorithm using tools from complexity
theory. 
If the unitaries describing the Hamiltonian are tensored Pauli operators, the number of measurements can be reduced substantially by employing shadow tomography~\cite{huang2020predicting,aaronson2020shadow}.
\end{document}